\documentclass{elsart}

\usepackage{rotating}

\usepackage{natbib}
\usepackage{graphicx}
 
\newcommand{\be}{\begin{equation}}
\newcommand{\ee}{\end{equation}}

\newcommand{\ba}{\begin{eqnarray}}
\newcommand{\ea}{\end{eqnarray}}

\newcommand\eg{\textit{e.g.\ }}

\newcommand{\Bf}{{magnetic field\,}}
\newcommand{\Bfs}{{magnetic fields\,}}

\begin{document}
\begin{frontmatter}

\title{Magnetic fields in GRBs}
\author{Maxim Lyutikov}

\begin{abstract}
We discuss the dynamical role of \Bfs  in GRB explosions in the framework of the fireball
 and electromagnetic models. We contrast the observational predictions of the two models and 
argue that only the very early afterglow observations, almost coincident with the prompt phase,
can be used as decisive tests of the  ejecta content.
\end{abstract}

\end{frontmatter}

\section{Introduction}

One of the key issues in GRBs physics is whether \Bfs  play an important dynamical role
at any stage in the outflow. Currently, the overwhelming point of view,
 advocated by the fireball model (FBM)
 is that
\Bfs do not play any major dynamical role. 
In the frame work of the FBM \Bf are 
 small scale, with correlation length $l_c$ much smaller
than the "horizon" length $l_c \ll R/\Gamma$ ($R$ is the radius of the outflow in the 
laboratory frame and $\Gamma$ is a bulk Lorentz factor).
 Small scale \Bfs can be created locally
in the flow; once created
 their energy density falls off faster with radius than plasma energy density, so that they
 do not
affect the overall dynamics of the outflow. 

Alternative approach, advocated by MHD and electromagnetic models \cite[\eg][]{lb03}
is that there are dynamically important
 large scale fields with "super-horizon" correlation length $l_c \geq R/\Gamma$.
Such \Bfs must be created at the source; 
at large distances the  toroidal component of the field dominates;
 it's energy density $B_\phi^2 \propto r^{-2}$ 
scales similar to plasma energy density $\rho c^2 \propto r^{-2}$ (for spherical outflow;
pressure falls off faster), so that they potentially can be important (or even dominate)
 for the overall dynamics of the flow. 
To quantify the dynamical
 importance of large scale \Bfs, it is useful to introduce magnetization parameter
$\sigma$ as a ratio of Poynting $F_{\rm Poynting} $ to (cold) particle $F_{\rm p}$  fluxes
\be 
\sigma= {F_{\rm Poynting} \over F_{\rm p} } = 
{B^2 \over 4 \pi \Gamma \rho c^2} = {b^{\prime 2} \over 8 \pi \rho ' c^2}
\ee
where $B$ and $\rho$ are \Bf and plasma density  in the lab frame, $b^{\prime}$ and $\rho '$
are \Bf and plasma density  in the plasma frame (where electric field is zero).
In the  FBM 
  $\sigma \ll 1$, MHD models work in the regime $\sigma \sim 1$,
EMM model assumes $\sigma \gg 1$. The question of the GRB model is then reduced to the question how 
large is $\sigma$ in the ejecta? 

The current state of the GRB theory is such that too often 
all the results are interpreted in terms of FBM,
 so that any new discovery is seen as a confirmation of it, while alternative
models are often called an extension of a FBM.
 In an effort to clear
up the terminology
 we suggest the following definitions based on the energy content of the GRB ejecta 
(due to page constraints we limit our discussion to two models: 
fireball and electromagnetic). 
These two types of models are at the extreme limits of $\sigma$; one may imaging an intermediate, 
MHD  models with $\sigma \sim 1$. 

{\bf Fireball model \cite[FBM, \eg][]{Piran04}}: 
 The defining characteristic is that most  energy produced by the central source  is 
carried by the bulk motion of  ions. In  temporal order the transformations of energy are as follows.
 The initial source of energy is not specified. Initial energy is thermalized near the central source,
 so that most of it
is converted into lepto-photonic plasma.
 This internal energy is then converted to the bulk motion of  ions,
 and reconverted back into internal at internal shocks; at the same time, the
 small scale \Bfs are generated.
The energy of these generated \Bfs is then used to accelerate leptons via Fermi mechanism 
to highly relativistic 
energies (note that the energy that goes to non-thermal particles is {\it electro-magnetic} 
even in the FBM:
Fermi-type acceleration is done by turbulent EMF associated with fluctuations of \Bf). 

{\bf Electro-magnetic model, \cite[EMM, \eg][]{lb03}}. 
The defining characteristic of the electro-magnetic model is that the bulk energy of the flow
is carried by \Bf. In temporal order  the evolution of the energy proceeds as follows. 
The energy that will power a GRB comes from kinetic rotational energy of the central source
(millisecond pulsar or BH-disk system). It is then converted to magnetic energy using 
unipolar inductor (like in pulsars), transported to 
large distances and is used to accelerate particle directly through reconnection-type events. 

\section{Large scale \Bfs}

Theoretically, large scale, energetically dominant \Bfs are expected in the
launching regions of relativistic outflows.
This is exemplified by pulsars, which generate magnetized winds, and by AGN
jets, launched  and collimated by electromagnetic stresses
 (\eg through Blandford-Znajek mechanism). Latest full relativistic MHD
 numerical simulations of accretion on to black hole
do show formation of the strongly-magnetized axial funnel \cite{Gammie}.

\section{Look early!}
How can the two models be distinguished? 
Below we argue that only direct observations at the prompt phase and very early afterglows  (AG)
can be used to distinguish the model: 
 observations of the late afterglows can  be used
 only as hints at best.

AG evolution may be separated into two phases
 with very different dynamics which we will call  early and late AG.\\
{\bf I. Early AG}.
In a  FBM, the  early AG phase is when the  total swept-up mass is of the order of the eject mass
$t \sim (E_\Omega/ \Gamma_0^2 \rho_{ISM} c^5)^{1/3}$ where $E_\Omega$ is the 
 energy of explosion per unit solid angle, 
$\Gamma_0$ is initial Lorentz factor and
$t$ is coordinate time (we assume $ \rho_{ISM}= constant$). At the early AG
 the Lorentz factor
of the flow is constant. 
In EMM, the  early AG is when the fast magneto-sonic waves emitted by the source
are still catching with the surface of the expanding bubble.  This stage ends at  
$t\sim   (E_\Omega t_s/  \Omega \Gamma_0^2 \rho_{ISM} c^5)^{1/4} $ where 
$t_s \sim 30 $ s is the source activity period (in observer time $t_{ob}$ this stage ends
when $t_{ob} \sim t_s$).
  At the early AG  the  Lorentz factor
is decreasing $\Gamma \propto t^{-1/2}$ and is much higher than in the FBM. \\
{\bf II. Late AG}.
At the late AG stage, when most of the energy has been transferred to the forward shock, 
the system forgets (almost completely, see below) what was the content of the ejecta (\eg ions or \Bf);
this stage is approximated by  Blandford-McKee self-similar solution
(\eg, for constant outside density and in the non-radiative regime $\Gamma \propto r^{-3/2}$).
{\it 
The overall
 temporal behavior of late AG is  determined by the total energy release (and not the form
of that energy) and, as a consequence, can hardly be used to distinguish between the models.
On the other hand, early AG, which are virtually coincident 
with the prompt phase, have very different
  behavior and can be used to distinguish between the models.}

There are still {\it  subtle } properties 
 of late AGs that may   serve as indications
in favor of one or the other model.
Beside the explosion energy,   relativistic
 forward shock still "remembers" the angular  distribution of the
deposited energy $E(\theta)$. 
This "memory" comes, first, from the fact that two points on the shock wave front
separated by the angle $\Delta \theta \geq 1/\Gamma$ are causally disconnected, and secondly,
from  relativistic kinematics
  freeze-out of forces normal to the shock front
\cite[\eg][]{sha79} even inside the cone  $\Delta \theta \leq  1/\Gamma$. 
As a result,  $E(\theta) $ remains approximately constant
 in time (ballistic expansion) until  $\Gamma \sim $ several, after which point  
 the non-spherical forward shock starts to evolve laterally toward sphericity. 
\footnote{
We would like to stress that 
 at the late AG stage there no "jet", but a non-spherical shock wave  evolving  in time.
 The expressions like  "jet break times",
"jet sideways expansion" etc are all misnomers.  
}

Thus, {\it 
the angular  distribution of the total energy $E(\theta)$ can 
be used to distinguish between different models, if a model predicts it}.  
This can be related to the
 extensive discussion about the structure of the GRB outflows
("uniform jet" or "structured  jet"),
or, to be more precise, about the structure of the forward shock generated by the explosion.
The implications of one or the other profile  seem  to be never mentioned.
The 
FBM  does not make any predictions for $E(\theta)$.
 The electromagnetic model, on the other  hand, does predict $E(\theta)\propto1/\theta^2$, 
the  "structured  jet".

 Another possible  hint from later AG  may come from polarization measurements. 
If magnetic field is produced locally, the polarization fraction and the position angle
 must correlate with
 "jet break time" (for "uniform jets" polarization fraction goes through zero while
 the  position angle experiences a $90^\circ$ flip at the "jet break time"; for "structured
jets" the  polarization fraction peaks at the moment
of "jet break" while the position angle remains constant).
 In the EMM,  it is feasible that  ejecta magnetic field mixes with the shocked plasma
and provides a source of polarization.  In this case  the position angle 
remains constant through the AG,  while the polarization fraction is constant before 
the "jet break" and decreases afterward \cite{lazz04}.

\section{Tests of GRB models}

{\it
 Decisive tests of different GRB models are bound to come from  observations  (almost) 
contemporaneous 
with the prompt emission}. 

{\bf Reverse shok emission}.
Perhaps the simplest test of GRB models may come from observations of emission from the  reverse
shock propagating in the ejecta,  which typically falls
into the optical range. 
Note, that such a key observation that
may validate  a GRB theory may come from a small scale experiment, like ROTSE or RAPTOR.
FBM predicts (modulo the  adjustable parameters like the
 ratio of magnetic to plasma energy densities)
strong reverse shock emission, so that
absence of nearly contemporaneous optical emission in most GRBs is a strong argument against
FBM (MHD modeling of reverse shock emission seem to favor high $\sigma$ in the ejecta, 
\cite{zhang04}).
 In EMM reverse shock is absent (optical flash may be produced by other mechanisms, like 
gamma-ray pair production in front of the forward shock, \cite{belob}).

\begin{figure}
\includegraphics[width=0.95\linewidth]{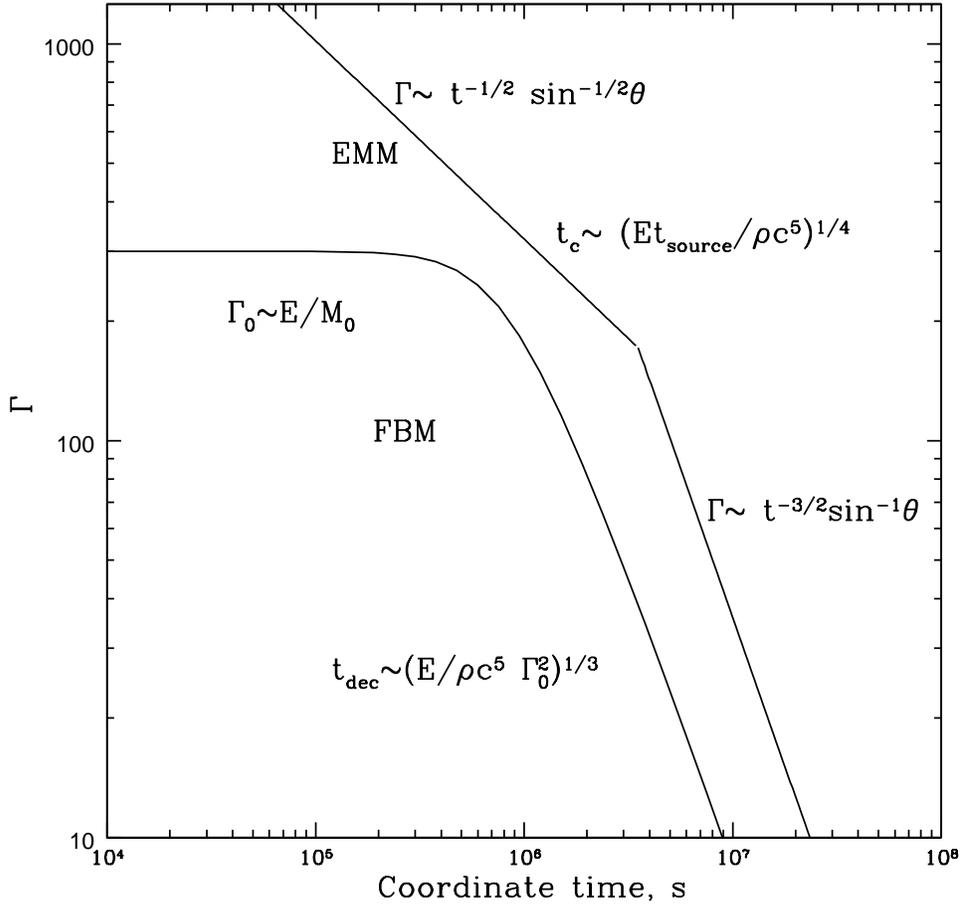} 
\caption{Evolution of Lorentz factors in the FBM and EMM at  early AG
$\,t_{ob} \leq t_{GRB}$ for constant external density.
Overall normalization of the curves depends
on the viewing angle (in the EMM).
Generally, at a given time the Lorentz factors in EMM are higher than in the FBM.}
\label{fig}
\end{figure}

{\bf Spectra of early AGs}.
FBM and EMM make very different prediction for the properties of early AGs (see Fig \ref{fig}).
According to EMM, at the early AG stage the Lorentz factor and peak frequency are
  much larger (and falling with time) than in the
FBM (constant  Lorentz factor and  peak frequency). Thus, early afterglow in the EMM
are more energetic than in FBM and can blend with the prompt phase.
 Practically, the 
  observations required to distinguish the models
 are tricky, both because of the tight temporal constraints, and because the
 early AG emission is faint: it increases as the burst progresses, reaching maximum 
at the end of prompt/early AG phase.

 \renewcommand{\baselinestretch}{1.5}
{\Large
\begin{sidewaystable}[p]
\begin{tabular} {l|ll|ll}
&\hskip .5 truein FBM \hskip -.3 truein & &\hskip .5 truein  EMM  \hskip -.3 truein  &\\
& $\rho = const$  & $\rho= \rho_0 (r_0/r)^2$ &  $\rho = const$  & $\rho= \rho_0 (r_0/r)^2$ \\
\hline
Lorentz factor $\Gamma$ & ${E_0 c^2 \over \dot{M}_0} $  & 
 ${E_0 c^2 \over \dot{M}_0} $   &  $ \left({ L \over \rho c^5  }\right)^{1/8}  \,t_{ob}^{-1/4 } $ &
 $\left({L \over \rho_0 r_0^2 c^3 }\right)^{1/4} $ \\
Peak frequency $\nu_m$ & $ {m_p^2 e \sqrt{\rho} \over m_e^3}
 \Gamma_0 ^4 \epsilon_e^2 \epsilon_B^{1/2}$ &
$ \sqrt{ 2 \pi} { m_p^2 e \sqrt{\rho_0} \over m_e^3 c} r_0 \Gamma_0 ^2 \epsilon_e^2 \epsilon_B^{1/2}  t_{ob}^{-1} $ & 
$  {m_p^2 e \sqrt{L } \over m_e^3 c^{5/2}} \epsilon_e^2 \epsilon_B^{1/2}   \,t_{ob}^{-1}  $ & 
$  {m_p^2 e \sqrt{L } \over m_e^3 c^{5/2}} \epsilon_e^2 \epsilon_B^{1/2}   \,t_{ob}^{-1}  $  \\
Peak flux $F_{nu, {\rm max}}$ & ${ e^3 c^2 \Gamma_0^8 \rho^{3/2} \over D^2  m_e^2 m_p} \epsilon_B^{1/2} \,t_{ob}^2 $ &
 $  { e^3 \Gamma_0^2 r_0^3 \rho_0^{3/2} \over m_e^2 m_p c D^2 } \epsilon_B^{1/2}$ &
$ { e^3 L  \sqrt{\epsilon_B} \sqrt{\rho} \over c^3 D^2  m_e^2 m_p}  \,t_{ob} $ &
$ { e^3 \sqrt{L} r_0^2 \rho_0 \over c^{5/2} m_e^2 m_p D^2 } \epsilon_B^{1/2}  $  \\
Cooling frequency $\nu_c$ &  $   { 1\over 64 \sqrt{2} \pi^{3/2}} { m_e ^5 c^6  \over e^7 
\Gamma_0^4 \epsilon_B^{3/2} \rho^{3/2} }  \,t_{ob}^{-2}$ & 
$ { c^9 m_e^5 \Gamma_0^2 \over 16 \sqrt{2} \pi^{3/2} e^7 r_0^3 \epsilon_B^{3/2}  } t_{ob} $
 & $  { m_e ^ 5 c^{17/2} \over
e^7 \sqrt{L}  \rho \epsilon_B^{3/2} }  \,t_{ob}^{-1}  $ & 
${ c^{15/2} m_e^5  \sqrt{L} \over  8 \sqrt{2} \pi^{3/2} r_0^4  \epsilon_B^{3/2} \rho_0^2} t_{ob} $ \\
End  of early AG, $\,t_{ob}$  & $ \left( { E_\Omega \over \Gamma_0^8  \rho c^5} \right)^{1/3} $&
${  E_\Omega \over 2 c^3 r_0^2 \Gamma_0^4 \rho_0} $  &$ t_s \sim t_{GRB}$ & $  t_s \sim t_{GRB}$ 
\end{tabular}
\caption{Properties of early afterglows. Post-shock plasma is parametrized by $\epsilon_e$, the
ratio of electron energy density to total energy density, and $\epsilon_B$, 
the ratio of \Bf energy density to total. $D$ is distance to the source. }
\end{sidewaystable}
}

 \renewcommand{\baselinestretch}{1}

{\bf Polarization of prompt emission}.
Claims of high polarization \cite{coburn03}, if confirmed, 
may provide a decisive test of GRB models.   
There is no question that the best way to 
produce polarization in the limit $10\% \leq \Pi \leq 60\%$
is through synchrotron emission in  large scale \Bfs \cite{lyu03c}.
(Larger polarization can only be produced with inverse Compton mechanism, 
smaller  polarization can be produced by small scale \Bfs.)
The FBM need to make several independent assumptions and fine tuning of parameters
in order to produce high polarization:
(i) jet opening angle $\sim 1/\Gamma$, 
(ii) viewing angle  $\sim 1/\Gamma$, (iii)
small scatter of Lorentz factors (this runs
 contrary to the very basic assumption of the FBM!), 
(iv) nearly perfect alignment of all  shocks' normals with radial direction
{\it in the  bulk frame},  (v) two-dimensional turbulent \Bfs. 
In order to get polarization in double digits the FBM needs to fine tune all these
factors independently.

{\bf 
Observational consequences of the EMM}
Observational properties of the EMM listed below follow from the assumption that the central
source generates  strongly
 magnetized wind with small baryon loading,
 and that the electrical current in the wind is concentrated near the 
axis:
\begin{itemize}
\item Weak thermal precursor. If a fraction $1/\sigma\sim 0.01-0.1$
of the magnetic energy is dissipated near the source, this should produce
a thermal precursor with luminosity $\sim 0.01-0.1$ of the
main GRB burst. FBM generally predict strong thermal precursor
\cite[\eg][]{dgaine}.
\item High polarization of prompt emission.
\item  Constant position angle of AG polarization along the projection of the
axis on the plane of the sky (same  position angle 
as in prompt if both are seen)
\item Hard-to-soft 
spectral evolution (similar to "radius-to-frequency" mapping in pulsars). 
At later times emission is generated at larger radii, where \Bf is lower.
\item $\epsilon _{peak} \propto \sqrt{L_\Omega}$. 
In EMM, $B \sim \sqrt{L_\Omega}$  (since $L \sim B^2 r^2 c$);
for synchrotron emission, the  peak energy is proportional to magnetic field.
\item No emission from the reverse shock. Strongly magnetized 
plasma supports only weak shocks, if at all. 
\item Decreasing peak frequency during early AG (for constant density
surrounding). 
\item "Universal jet" structure of AG. Magnetically-dominated outflows
have a preferred distribution of currents: when most of the current is concentrated near the axis of the flow \cite{hava} (this configuration
corresponds to  a minimum energy given total magnetic flux). 
In this case $B \sim 1/\sin \theta$ and $L_\Omega \sim  1/\sin^2 \theta$.
\item AG peak flux  reaches its maximum  at the end of prompt phase (in FBM the moment of peak flux
is not tied to prompt emission duration).
\item "Jet break" observed at $\theta_{ob} \Gamma_{\rm axis} \sim 1$.
\item If AG is resolved at the "jet break" moment, 
it should appear distorted (dislocated) along the projection of the explosion
 axis on the plane of the sky,
 with apparent expansion velocities along the axis reaching 
$\sim 1/\theta_{ob}$; if polarized, position angle  should be along the axis
\item XRFs and GRBs form a continuous sequence.
\item Few orphan AGs.
\item  Same {total kinetic} energy in XRF and GRBs, since
$E_{\rm tot} \sim \int d \Omega L_\Omega \propto \ln \sin \theta_0$, 
where $\theta_0\sim 10^{-3}-  10^{-2}$ is the  angular size of the
current-carrying core.
\item  XRFs have on average the same duration as GRBs, determined by the 
activity period of the central source.
\end{itemize}

{\bf 
Theoretical implications of the EMM}
\begin{itemize}
\item Variability of the prompt emission reflects 
the statistical properties of dissipation
(and not the source activity as in the FBM). This is similar to the case
of solar flares, which show variability on a wide range of temporal scale unrelated to the time scale of \Bf generation in tachocline.
\item Variability may be enhanced  due to random relativistic motion of the
 "fundamental emitters"
in the outflow bulk frame. Highly magnetized medium can support disturbances
propagating with high Lorentz factor in the bulk frame; alternatively,
magnetic dissipation processes, like reconnection, may produce  
highly relativistic streams of plasma. 
\item ``Standard candle'' - the  narrow distribution
of GRB energies - may be related to 
critically rotating relativistic stellar mass object, a one
parameter family (under-energetic GRBs may be related to sub-critically rotating central 
objects)
\item Particle acceleration is due to dissipation of magnetic energy 
(and not shock acceleration
as in the FBM).  Investigation of magnetic 
dissipation in the strongly magnetized plasma
is only beginning \cite{lyut,Hosh}. 
\end{itemize}

The Swift mission, with its early response,
should be able to distinguish between FBM and EMM. 
At a basic lavel, Lorentz factors in the
EMM model are higher, which leads to  more energetic  early AGs 
with  higher peak frequencies.


\end{document}